\documentclass[aps,pra,twocolumn,superscriptaddress,showpacs]{revtex4-1}
\usepackage[utf8]{inputenc}
\usepackage[english]{babel}
\usepackage{amsmath}
\usepackage{amssymb}
\usepackage{graphicx}
\usepackage[colorlinks=true, urlcolor=blue, citecolor=blue, filecolor=blue, linkcolor=blue]{hyperref}
\usepackage{siunitx}
\usepackage{bigstrut}
\usepackage{tabularx}

\begin{document}

\title{Precision measurement of the rotational energy-level structure of the three-electron molecule He${_2}^+$}

\author{Luca Semeria}
\author{Paul Jansen}
\affiliation{Laboratory of Physical Chemistry, ETH Zurich, CH-8093 Zurich, Switzerland}
\author{Fr\'ed\'eric Merkt}
\email[]{merkt@xuv.phys.chem.ethz.ch}
\affiliation{Laboratory of Physical Chemistry, ETH Zurich, CH-8093 Zurich, Switzerland}

\date{\today}

\begin{abstract}
The term values of all rotational levels of the $^4$He${_2}^+\,X^+\,^2\Sigma_u^+\,(\nu^+=0)$ ground vibronic state with rotational quantum number $N^+\le 19$ have been determined with an accuracy of \num{8e-4}\,cm$^{-1}$ ($\sim{25}$\,MHz) by MQDT-assisted Rydberg spectroscopy of metastable He$_2^*$. Comparison of these term values with term values recently calculated \emph{ab initio} by Tung \emph{et al.} [\href{http://dx.doi.org/10.1063/1.3692800}{J. Chem. Phys. \textbf{136}, 104309 (2012)}] reveal discrepancies that rapidly increase with increasing rotational quantum number and reach values of \num{0.07}\,cm$^{-1}$ ($\sim{2.1}$\,GHz) at $N^+=19$.
\end{abstract}


\maketitle

\section{\label{sec:intro}Introduction}
Predicting the energy level structure of molecules from first principles is one of the major goals and tasks of theoretical molecular physics and chemistry. Few-electron molecules such as H${_2}^+$, H$_2$ and He${_2}^+$ are particularly attractive systems for comparison with experimental results because numerically ``exact'' predictions of molecular properties can in principle be obtained, i.e., predictions that are only limited in accuracy by the uncertainties of fundamental constants. For H${_2}^+$ and H$_2$, remarkable progress has been achieved by theoretical calculations in the past ten years \cite{korobov2006,korobov2008,korobov2014-3,piszczatowski2009,komasa2011,puchalski2016}. The most precise theoretical calculations in these molecules are based on an effective nonrelativistic form of QED that treats relativistic and radiative effects perturbatively using the nonrelativistic wave functions. Remarkable progress has also been made in experimental determinations of energy intervals in these molecules \cite{osterwalder2004,koelemeij2007,liu2009,salumbides2011,dickenson2013,haase2015,biesheuvel2016} so that the prospect of exploiting the comparison of theoretical and experimental results in these systems to either improve the accuracy of fundamental constants \cite{wing1976} or uncover new physical phenomena \cite{schiller2005,ubachs2016} becomes increasingly attractive. For instance, Salumbides \emph{et al.} argue that precision measurements in molecules may reveal (or at least set upper bounds to potential effects of) so far unknown forces acting on nuclei at distances typical of molecular bonds \cite{salumbides2013,ubachs2016}. In general, precision measurements in simple systems are ideal to test theoretical predictions at the most fundamental level. The three-electron molecule He${_2}^+$, with nuclei four times as heavy as those in H${_2}^{(+)}$, is also an attractive molecular system in this context.

\emph{Ab initio} calculations in He${_2}^+$ were carried out almost immediately after Heitler and London showed how to efficiently treat electron correlation in H$_2$ \cite{heitler1927}. Majorana \cite{majorana1931} and Pauling \cite{pauling1933} extended the Heitler-London method to predict the dissociation energy, the equilibrium distance, and the harmonic vibrational frequency of He${_2}^+$. Since then, numerous calculations have been reported on He${_2}^+$ (see, e.g., Refs.~\cite{weinbaum1935,moiseiwitsch1956,csavinszky1959,reagan1963,liu1971,maas1976,khan1986,bauschlicher1989,cencek1995,cencek2000,xie2005,tung2012}). The most recent \emph{ab initio} calculations of the energy-level structure of He${_2}^+$ by Tung \emph{et al.} \cite{tung2012} provided almost exact nonrelativistic energies and an accuracy comparable to the accuracy of the best experimental results available on He${_2}^+$ in the literature, i.e., the 9 infrared rovibrational transition frequencies recorded between low-lying rovibrational levels in $^3$He$^4$He$^+$ by Yu and Wing \cite{yu1987} and the 7 rovibronic transition frequencies measured between high-lying vibrational levels of the $X^+\,^2\Sigma_u^+$ electronic ground state of $^4$He${_2}^+$ and the weakly bound levels of the $A^+\,^2\Sigma_g^+$ state near the He$^+$\,+\,He dissociation limit by Carrington and coworkers \cite{carrington1995}. Tung \emph{et al.} \cite{tung2012} reported the accuracy of their calculations to be limited by the neglect of relativistic and QED corrections, which they estimated to be on the order of 0.004\,cm$^{-1}$ (120\,MHz) for the low-lying rovibrational levels of He${_2}^+$ by comparing such corrections in H$_2$ with the relativistic correction calculated for the $\nu=0 - 1$ interval of $^3$He$^4$He$^+$ \cite{stanke2009}. 

Experimental measurements of the low-lying levels of $^4$He${_2}^+$ are complicated by the fact that $^4$He${_2}^+$ is apolar and has no electric-dipole-allowed vibrational and rotational transitions. Recently we have started systematic measurements of the energy-level structure of H${_2}^+$ and He${_2}^+$ by multichannel-quantum-defect-theory(MQDT)-assisted Rydberg spectroscopy of H$_2$ \cite{osterwalder2004,liu2009,sprecher2011,haase2015} and He$_2$ \cite{sprecher2014,jansen2015}. In the case of He${_2}^+$, the procedure consists of measuring high-resolution spectra of triplet ($S=1$), $n$p Rydberg states of very high principal quantum number $n$ from the $a\,^3\Sigma_u^+$ metastable state of He$_2$ and extrapolating the Rydberg series to their limits using MQDT \cite{jungen2011}. We report here the determination of the term values of all rotational levels of the $X^+\,^2\Sigma_u^+\,(\nu^+=0)$ state of He${_2}^+$ up to the rotational level with rotational quantum number $N^+=19$, which is located more than 2600\,cm$^{-1}$ above the ground state of He${_2}^+$. Surprisingly, we find that the deviations between measured and calculated term values rapidly increase with increasing $N^+$ value.

\section{Experimental procedure\label{sec:experiment}}
\begin{figure}[bt]
\centering
\includegraphics[width=1\columnwidth]{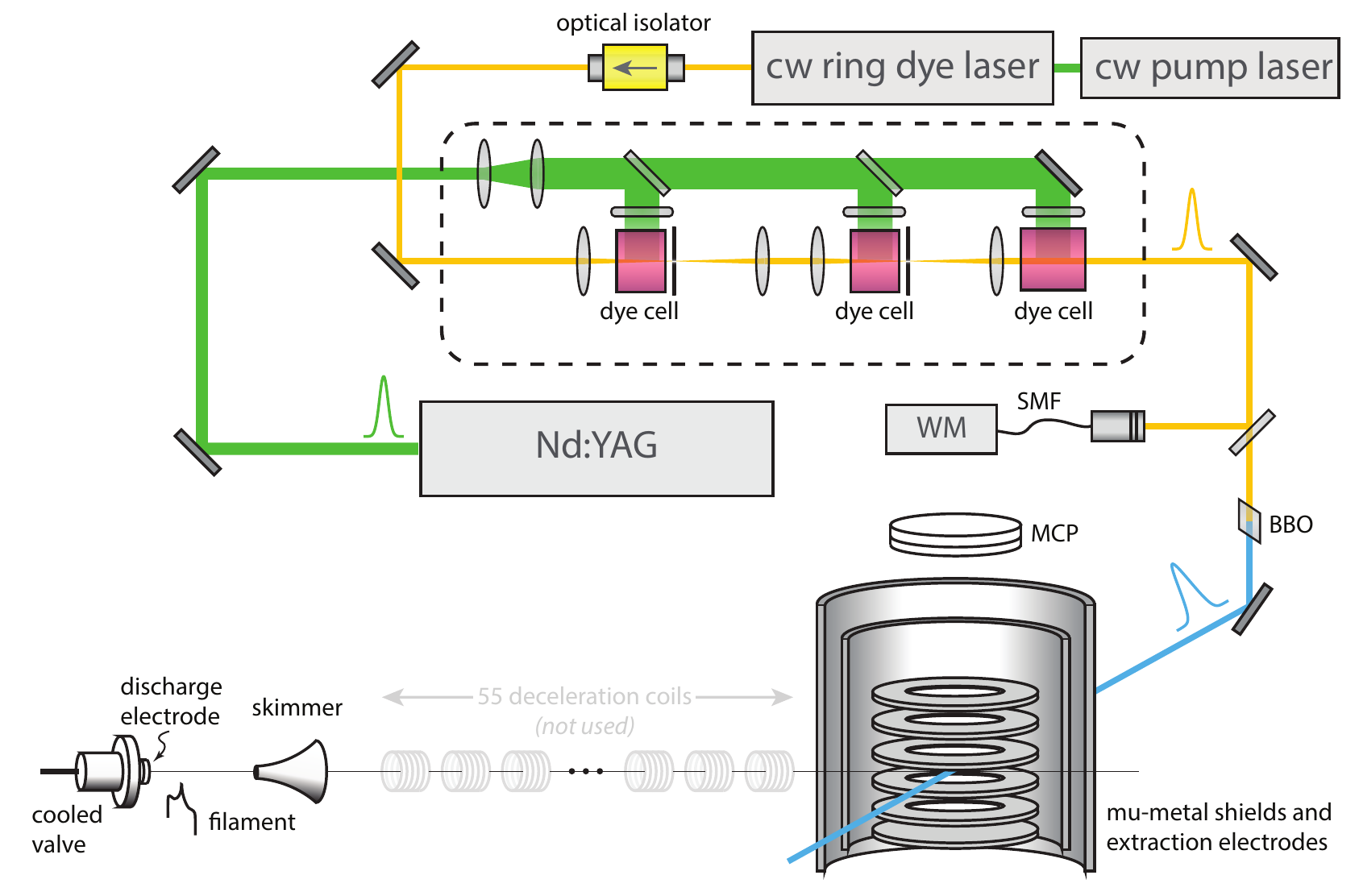}
\caption{Schematic representation (not to scale) of the experimental setup. Shown are the pulsed-amplified ring-dye-laser system and spectrometer containing the discharge source, the Zeeman decelerator, and the magnetically shielded photoexcitation region. Nd:YAG, neodymium-doped yttrium aluminum garnet; WM, wavemeter; SMF, single-mode fiber; BBO, beta-barium-borate crystal; MCP, microchannel-plate detector. 
\label{fig:setup}}
\end{figure}

The experimental setup is presented schematically in Fig.~\ref{fig:setup} (see also Refs.~\cite{jansen2015} and \cite{jansen2016}). The measurements were performed with a three-stage pulse-amplified laser system \cite{hollenstein2000} that was seeded with the output of a continuous-wave (cw) ring dye laser and pumped by the second harmonic of a Q-switched neodymium-doped yttrium-aluminium-garnet (Nd:YAG) laser. The pulse-amplified output of the ring laser was frequency doubled in a beta-barium-borate crystal, resulting in near-Fourier-transform-limited pulses of UV radiation with a bandwidth of 180\,MHz. The frequency was adjusted by scanning the seed-laser frequency. 

Calibration of the laser fundamental frequency to an accuracy of 20\,MHz (1$\sigma$) was achieved by coupling a fraction of the pulse-amplified output of the ring laser into a wavemeter (HighFinesse WS7). To quantify the effects of the frequency chirp arising in the pulse-amplification process, part of the cw output of the ring laser was used to record the laser-induced-fluorescence spectrum of I$_2$ so as to determine the difference between the frequencies of the cw and pulse-amplified outputs of the ring dye laser. 

A supersonic beam of metastable helium molecules was produced in a source chamber by striking an electric discharge through an expansion of pure helium gas at the exit of a pulsed valve \cite{raunhardt2008}. The body of the valve was cooled to a temperature of 77\,K, resulting in a supersonic beam with a velocity of approximately 1000\,m/s \cite{motsch2014}. The molecular beam was collimated with a 2-mm-diameter skimmer before entering a second, differentially-pumped vacuum chamber containing a 55-stage Zeeman decelerator \cite{vanhaecke2007,motsch2014,wiederkehr2011}. Although the Zeeman decelerator was not used to obtain the results presented below, it played an important role in the initial assignment of the spectra and in the assessment of the systematic uncertainty resulting from the Doppler effect, as explained in Refs.~\cite{jansen2015} and \cite{jansen2016}. 

After about 1\,m of flight, the molecules entered a third vacuum chamber that was used for photoexcitation and detection. The photoexcitation region was surrounded by a cylindrically symmetric stack of electrodes for the application of ionization and extraction electric fields. The molecular beam crossed the UV laser beam at right angles on the symmetry axis of the electrode stack, which was oriented perpendicularly to the plane defined by the laser and molecular beams. High-$n$ Rydberg states were ionized by a pulsed electric field, which was also used to extract the ions toward a microchannel-plate (MCP) detector. The spectra were obtained by monitoring the pulse-field-ionization yield as a function of the UV laser wavenumber. The spectra were greatly simplified by applying a small electric field shortly after photoexcitation but before the field-ionization pulse. This first ``discrimination'' pulse did not only field ionize Rydberg states with $n\gtrsim200$, but also shifted all ions produced by direct ionization or rapid autoionization (prompt ions) out of the photoexcitation volume. Consequently, the ions generated by the second, field-ionization pulse had different flight times to the MCP detector than the prompt ions and could be monitored separately. Stray electric fields were determined by recording spectra in the presence of different applied potentials across the stack and fitting the observed Stark shifts to a quadratic polynomial and were compensated to a level of better than 1\,mV/cm, as explained in Ref.~\cite{osterwalder1999}. To suppress stray magnetic fields, two concentric mu-metal tubes were used to shield the photoexcitation region. 

\section{Rydberg series of $\text{He}_2$ and Multichannel Quantum-defect Theory calculations\label{sec:mqdt}}
\begin{figure*}[tb]
\centering
\includegraphics[width=.75\textwidth]{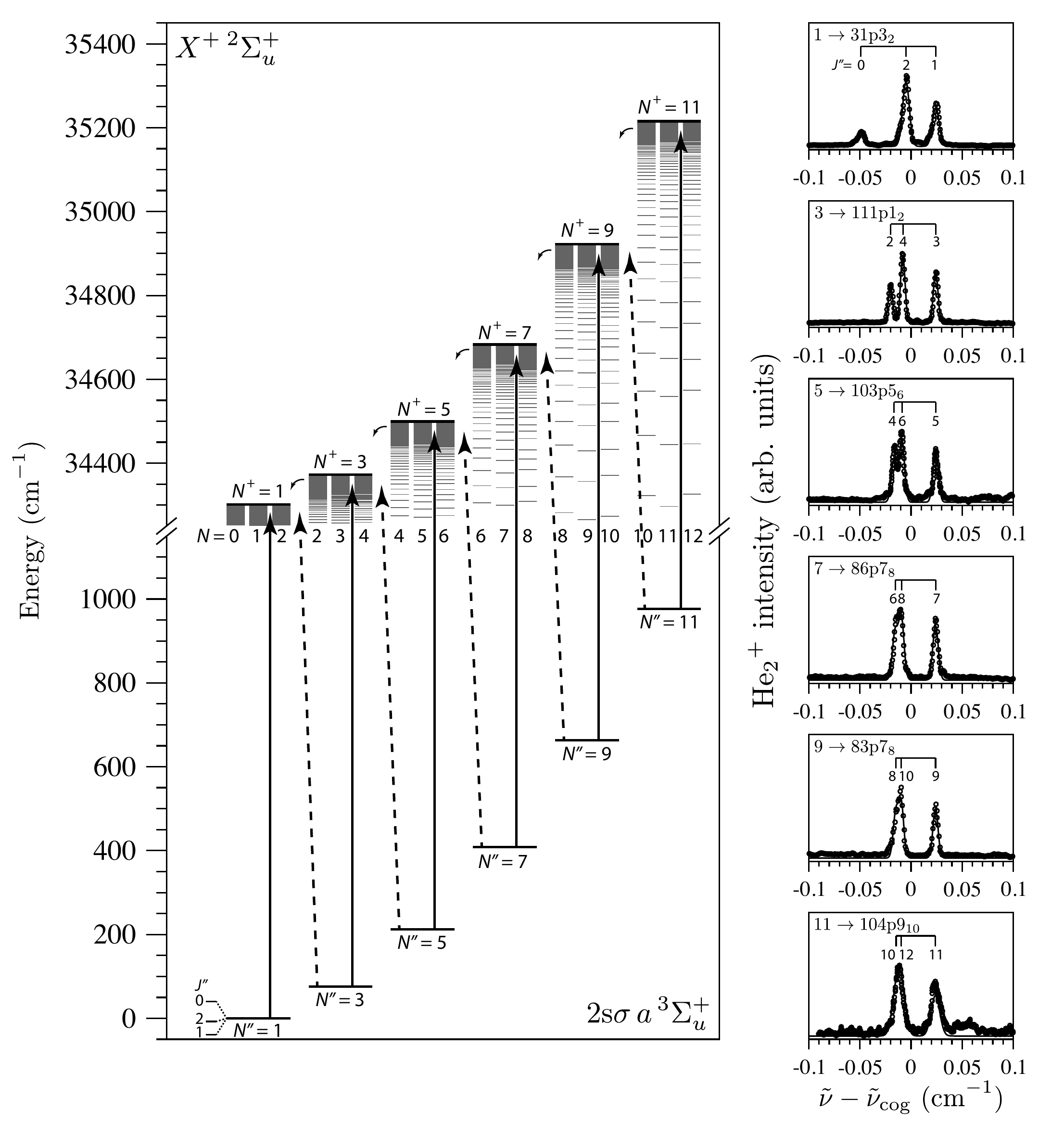}
\caption{Left panel: energy-level diagram showing the lowest six rotational levels of He$_2\,(a\,^3\Sigma_u^+,\,\nu''=0)$ and the triplet $n$p Rydberg series of He$_2$ converging on the lowest six rotational levels of He${_2}^+\,(X^+\,^2\Sigma_u^+,\,\nu^+=0)$. Full arrows represent the $n\text{p}N^+_N\leftarrow N''$ series with $N^+=N''$ and $N=N^+,N^++1$ that were used to determine the ionization thresholds, while dashed arrows indicate $n\text{p}N^+_N\leftarrow N''$ series with $N^+=N''-2$ and $N=N^++1$ that were used to determine the rotational structure of the $a\,^3\Sigma_u^+\,(\nu''=0)$ state of He$_2$. Curved arrows indicate the $n\text{p}N^+_{N=N^+-1}$ series that undergo rapid rotational autoionization above the $N^+-2$ threshold. The fine structure of the initial state is not visible on the scale of the figure and is shown strongly magnified for the $N''=1$ rotational state only. Right panels, top to bottom: Representative experimental spin-spin and spin-rotation fine structure  observed for the lowest six rotational states of He$_2^*$ (open circles) along with the fitted line profiles (full lines). The transitions are shown with respect to their fitted center of gravity (cog), with $\tilde{\nu}_\text{cog}=\text{\numlist{34258.51664;34216.44442;34277.23394;34260.14215;34003.79264;33934.84385}}$ cm$^{-1}$ from top to bottom.}
\label{fig:level_diagram}
\end{figure*}

\begin{figure*}[bht]
\centering
\includegraphics[width=.65\textwidth]{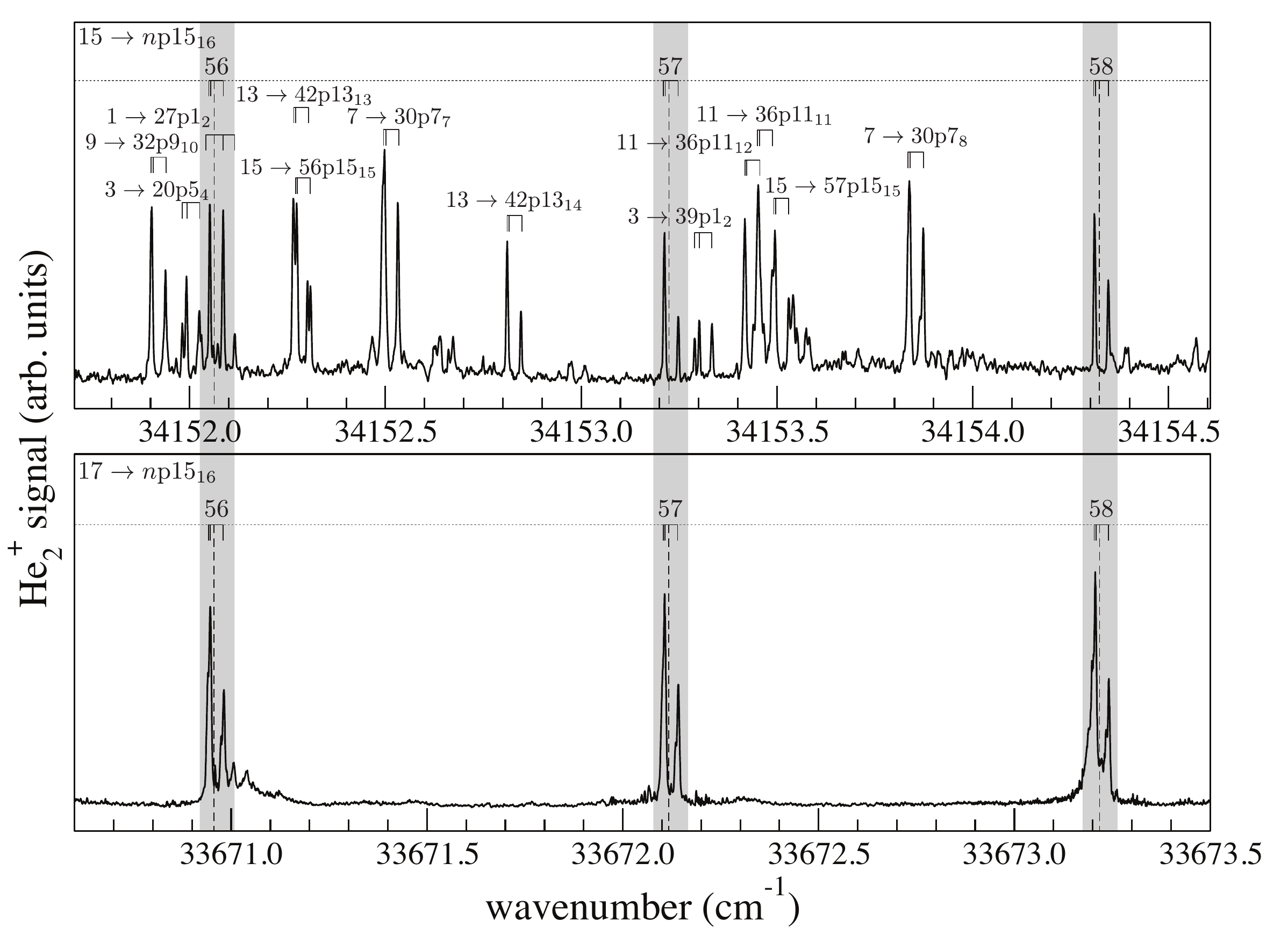}
\caption{Determination of initial-state combination differences with the example of the $n\text{p}15_{16}$ Rydberg series recorded from the $a\,^3\Sigma_u^+,\,\nu''=0,N''=15$ (upper panel) and $N''=17$ (lower panel) in the region $n=56-58$. The centers of gravity of the triplet structures are indicated by the vertical dashed lines. The background signal in the spectra is insignificant and the vertical scale is linear.}
\label{fig:Olines}
\end{figure*}

With the exception of the single bound level of the electronic ground state, all bound states of He$_2$ are Rydberg states~\cite{herzberg1987}. The rotational levels of the $a\,^3\Sigma_u^+$ metastable state and of high-$n$ Rydberg states are schematically depicted on the left-hand side of Fig.~\ref{fig:level_diagram}. The bosonic nature of the $^4$He nuclei excludes the existence of even rotational states in the $a\,^3\Sigma_u^+\,(\nu''=0)$ state of the neutral molecule and in the $X^+\,^2\Sigma_u^+\,(\nu^+=0)$ state of the ion, so that half of the energy levels is missing. In the following, we use double-primed symbols to indicate the quantum numbers of the metastable He$_2$ molecule and symbols with a superscript ``$+$'' to indicate the quantum numbers of the molecular ion. Single-photon excitation from the metastable $a\,^3\Sigma_u^+$ state, which is a triplet ($S''=1$) Rydberg state of s character \cite{ginter1965}, primarily results in the excitation of p Rydberg series converging on the different rotational levels of the $X^+\,^2\Sigma_u^+$ ground state of He${_2}^+$. The spin-rotation splittings of the rotational levels of the $X^+\,^2\Sigma_u^+$ ground state of He${_2}^+$ are not resolved in our spectra, so that the ionic rotational levels are adequately labeled by the rotational quantum number $N^+$. The total angular momentum (excluding spins) $\vec{N}$ of the high Rydberg states results from the vector addition of $\vec{N}^+$ with the Rydberg-electron orbital angular momentum $\vec{\ell}$ so that $N=N^++\ell,N^++\ell-1,\ldots,|N^+-\ell|$. For the p Rydberg states of interest here, this results in $N=N^+,N^+\pm 1$ (see Ref. \cite{raunhardt2008} for details). Consequently, the Rydberg states can be labeled as $n\text{p}N^+_N$. Each rotational level of the $a\,^3\Sigma_u^+$ metastable state is split into a triplet by the spin-spin and spin-rotation interactions \cite{lichten1974}. The rotational levels are labeled by the rotational quantum number $N''$ and the total angular momentum quantum number $J''$ ($J''=N'',N''\pm 1$). Each rotational level consists of a triplet with characteristic splittings, which are known extremely precisely from the high-resolution microwave studies of Lichten \emph{et al.} \cite{lichten1974,lichten1978} and Bjerre \emph{et al.} \cite{kristensen1990,hazell1995}. Consequently, all transitions have a triplet structure with characteristic splittings and intensity patterns that can be used to assign the initial level in a straightforward manner. Examples of these patterns for transitions originating from the $N''=1-11$ levels are given on the right-hand side of Fig.~\ref{fig:level_diagram}.  Three Gaussian functions of full width at half maximum of 180\,MHz, separated by the known fine-structure intervals \cite{lichten1974,lichten1978,kristensen1990,hazell1995,focsa1998} and with relative intensities weighted by the $2J''+1$ degeneracy factors, were used to convert each observed triplet into its ``fine-structure-free'' center of gravity. All transition wavenumbers reported in this article correspond to such centers of gravity.

Rydberg series converging on a particular ($\nu^+,N^+$) rovibrational level of the $X^+\,^2\Sigma_u^+$ electronic ground state of He${_2}^+$ are approximately described by Rydberg's formula

\begin{equation}
\tilde{\nu}_{n\ell}=\tilde{\nu}_{N'',N^+}-\frac{\mathcal{R}_{\text{He}_2}}{\left ( n-\delta_\ell\right )^2},\text{ with }hc\tilde{\nu}_{N'',N^+}=E_\text{I}+E_\text{rv}^+-E_\text{rv}'',
\end{equation}
where $\tilde{\nu}_{n\ell}$ represents the spectral position of the Rydberg states of principal and orbital angular-momentum quantum numbers $n$ and $\ell$, respectively, and quantum defect $\delta_\ell$. The quantities $\mathcal{R}_{\text{He}_2}$ and $E_{\text{I}}$ represent the mass-corrected Rydberg constant for He$_2$ and the adiabatic ionization energy of metastable He$_2$, corresponding to the energy difference between the $X^+\,^2\Sigma_u^+\,(\nu^+=0,N^+=1)$ level of He${_2}^+$ and the $a\,^3\Sigma_u^+\,(\nu''=0,N''=1)$ metastable level, respectively. $E_\text{rv}^+$ and $E_\text{rv}''$ are the rovibrational energies of the He${_2}^+$ ion core and of metastable He$_2$ with respect to the ($\nu^+=0,N^+=1$) and ($\nu''=0,N''=1$) levels, respectively. Although Rydberg's formula accurately describes the essentially unperturbed $n\text{p}N^+_{N=N^+}$ series, it does not account for the interactions between series of the same $N$ value that converge on different rotational states of He${_2}^+$. For the $n\text{p}[N^+=N+1]_N$ series, these interactions give rise to spectral perturbations below, and to rotational autoionization above, the $N^+=N-1$ thresholds (see Fig.~\ref{fig:level_diagram}) and can be accounted for in the framework of multichannel quantum-defect theory \cite[and references therein]{jungen2011}.

In this work, we determined the ionization thresholds corresponding to the different rotational levels of He${_2}^+$ with respect to the $N''=1$ rotational state of He$_2^*$ by extrapolating Rydberg series of metastable helium molecules using the MQDT program developed by Jungen \cite{jungen2011}. The data used for the extrapolation consist of $n$p Rydberg series with $n$ ranging from 95 to 110, excited from the $\nu''=0$ and $N''\le 19$ levels of metastable He$_2$. 

The quantum-defect parameters for the \emph{gerade} triplet $n$p and $n$f Rydberg states of He$_2$, derived by Sprecher \emph{et al.}~\cite{sprecher2014}, was adopted without change for the extrapolation of the Rydberg series. In Ref.~\cite{sprecher2014}, internuclear-distance($R$)-dependent and energy($\epsilon$)-dependent $\eta$ eigenquantum-defect matrices were extracted from data available in the literature on $n$p and $n$f Rydberg levels with $n$ in the range $2-39$, $v^+\le 2$, and $N^+\le 19$ in a least-squares-fit procedure. High-$n$ Rydberg states were not used in the fit, because the level energies of high-$n$ Rydberg states are insensitive to adjustments of the quantum defects, a fact that is also exploited in the present analysis (see Ref.~\cite{sprecher2014} for details). 

\begingroup
\begin{table*}[bt]
 \caption{\label{tab:compare_focsa} Rotational term values of the $a\,^3\Sigma_u^+\,(\nu''=0)$ state of He$_2$ (3rd column) and the $X^+\,^2\Sigma_u^+\,(\nu^+=0)$ ground state of He${_2}^+$ (6th column). For comparison, term values calculated with the molecular constants for the $a\,^3\Sigma_u^+\,(\nu''=0)$ state of He$_2$ reported by Focsa \emph{et al.} \cite{focsa1998} and the term values of the $X^+\,^2\Sigma_u^+\,(\nu^+=0)$ state of $^4$He${_2}^+$ calculated \emph{ab initio} by Tung \emph{et al.} \cite{tung2012}, are shown in the 2nd and 5th columns respectively. All values are given in cm$^{-1}$.}

\begin{tabularx}{.75\textwidth}{@{\extracolsep{\fill}}l S S c l S S@{}}
\toprule
\\[-1.8ex]
$N''$  &  \multicolumn{1}{c}{Ref.~\cite{focsa1998}\footnote{Calculated using the rotational constants reported by Focsa \emph{et al.} \cite{focsa1998}.}} & \multicolumn{1}{c}{This work} & & $N^+$ & \multicolumn{1}{c}{Ref.~\cite{tung2012}}   & \multicolumn{1}{c}{This work\footnote{The observed rotational term values can be described by $B_0^+=\num{7.101143(25)}$\,cm$^{-1}$, $D_0^+=\num{5.2156(36)e-4}$\,cm$^{-1}$, $H_0^+=\num{3.29(16)e-8}$\,cm$^{-1}$, and $L_0^+=\num{-7.3(22)e-12}$\,cm$^{-1}$}} \\
   & \multicolumn{1}{c}{(FTIR)}             &     & & & \multicolumn{1}{c}{(\emph{ab initio})} & \multicolumn{1}{c}{(MQDT fit)}\\\cline{1-3}\cline{5-7}
\\[-1.8ex]
 1   &     0                 &     0                 &  &  1  &     0      &    0                \\
 3   &    75.8129\pm0.0003   &    75.8136\pm0.0008   &  &  3  &    70.936  &   70.9379\pm0.0008  \\
 5   &   211.9937\pm0.0008   &   211.9950\pm0.0008   &  &  5  &   198.359  &  198.3647\pm0.0008  \\
 7   &   408.0605\pm0.0016   &   408.0614\pm0.0008   &  &  7  &   381.822  &  381.8346\pm0.0008  \\
 9   &   663.322\pm0.003     &   663.3231\pm0.0008   &  &  9  &   620.683  &  620.7021\pm0.0009  \\
11   &   976.879\pm0.004     &   976.8809\pm0.0008   &  &  11 &   914.112  &  914.1367\pm0.0008  \\
13   &  1347.638\pm0.007     &  1347.6396\pm0.0008   &  &  13 &  1261.089  & 1261.1242\pm0.0008  \\
15   &  1774.307\pm0.011     &  1774.3072\pm0.0008   &  &  15 &  1660.420  & 1660.4627\pm0.0009  \\
17   &  2255.413\pm0.017     &  2255.4133\pm0.0008   &  &  17 &  2110.736  & 2110.7932\pm0.0009  \\
19   &  2789.31\pm0.03       &  2789.3056\pm0.0008   &  &  19 &  2610.505  & 2610.5744\pm0.0009  \\
\toprule
\end{tabularx}
\end{table*}
\endgroup

\section{Data analysis and results\label{sec:results}}
\begin{figure*}[thb]
\centering
\includegraphics[width=.9\textwidth]{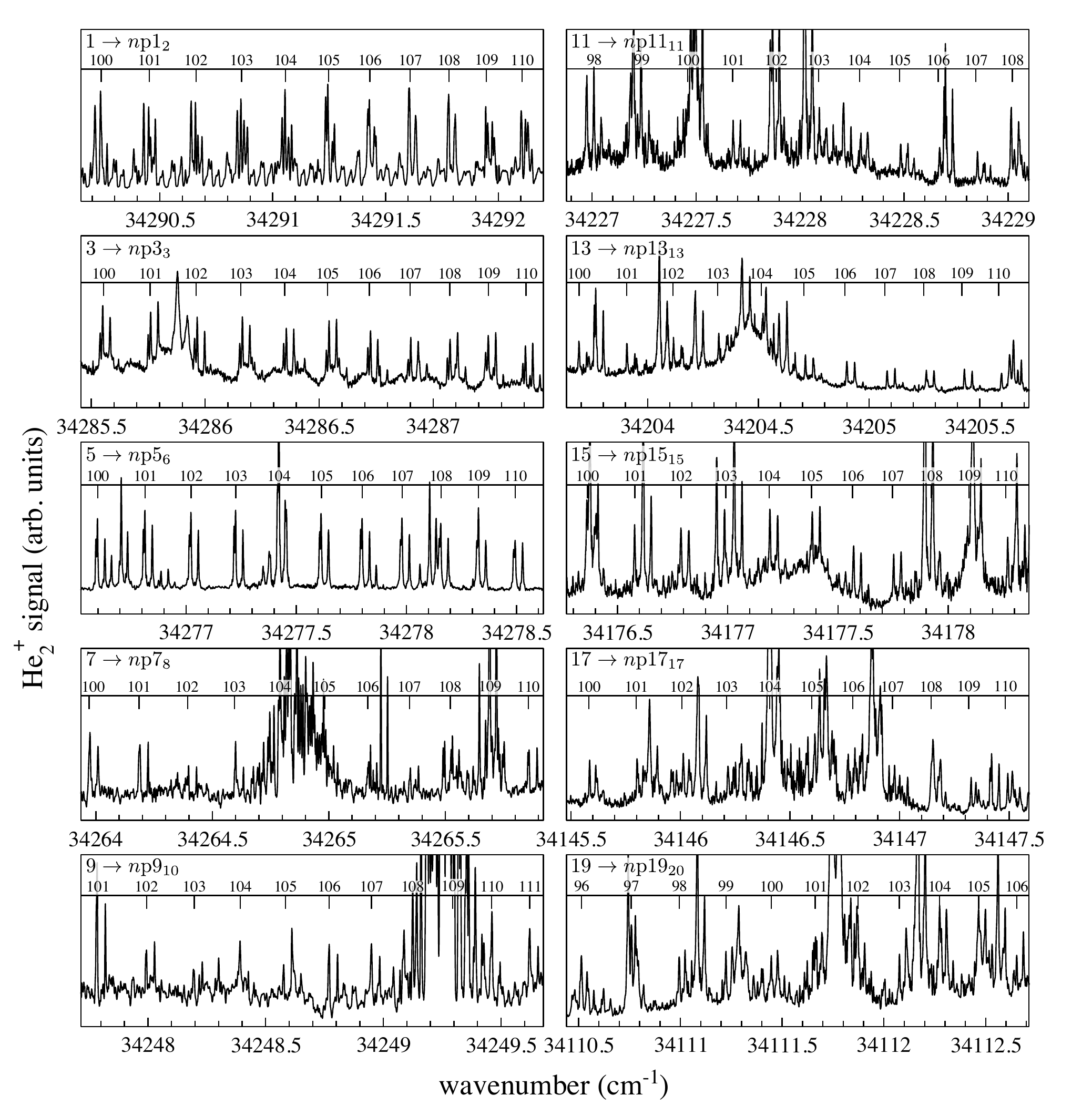}
\caption{Spectra of the $n\text{p}N^+_N\leftarrow a\,^3\Sigma_u^+\,(\nu''=0,N'')$ series of He$_2$ with $N^+=N''\leq 19$, $N=N^+,N^++1$ and $n$ around 100. For clarity, assignment bars are only shown for a single series in each spectrum. The background signal in the spectra is insignificant and the vertical scale is linear.}
\label{fig:Qseries}
\end{figure*}

The dominant $n\text{p}N^+\leftarrow a\,^3\Sigma_u^+\,(\nu''=0,N'')$ Rydberg series in the spectrum of He$_2$ are those with $N^+=N''$. There are three such series with $N=N^+,N^+\pm1$. However, all $n\text{p}N^+_{N^+-1}$ Rydberg states with $N^+>1$ located above the $N^+-2$ ionization thresholds are subject to rapid autoionization, because of the strong coupling, indicated by the curved arrows in Fig.~\ref{fig:level_diagram}, between channels of the same value of $N$, but differing in $N^+$ by 2 \cite{raunhardt2008,sprecher2014}. The corresponding transitions are strongly broadened and thus not detectable in our high-resolution spectra. Below the $N^+-2$ ionization thresholds, the autoionization of the $n\text{p}N^+_{N^+-1}$ Rydberg states is suppressed but the coupling between the $n\text{p}N^+_{N^+-1}$ and $n\text{p}[N^+-2]_{N^+-1}$ channels leads to strong perturbations and mixing of the Rydberg series. The $n\text{p}N^+_{N^+-1}$ series is always observable from both the $N''=N^+$ (full arrows in Fig.~\ref{fig:level_diagram}) and $N''=N^+-2$ (dashed arrows in Fig.~\ref{fig:level_diagram}) rotational levels of the metastable state, which enables one to determine the relative position of the $N''-2$ and $N''$ levels by building combination differences.

\begingroup
\squeezetable
\begin{table*}[t!]
\caption{\label{tab:Qlines}
Observed transitions from the $a\,^3\Sigma_u^+$ $(\nu''=0, N'')$ state of $^4$He$_2$ to the $n$p$N^+_N$ Rydberg states belonging to series converging to the $X^+\,^2\Sigma_u^+$ $(\nu^+=0, N^+)$ states of $^4$He${_2}^+$ and comparison with the results of MQDT calculations. The symbol $\Delta$ represents the difference between observed and calculated line positions ($\Delta=\text{obs.}-\text{calc.}$). All values are given in cm$^{-1}$. Missing entries correspond to transitions that were not observed or to strongly blended transitions.}
\begin{tabular}{
s[table-unit-alignment = left,table-format=3]
S[table-format=6.5]
S[table-format=-2.5]
c
S[table-format=6.5]
S[table-format=-2.5]
c
S[table-format=6.5]
S[table-format=-2.5]
c
S[table-format=6.5]
S[table-format=-2.5]}

\hline\hline
& \multicolumn{2}{c}{$1\rightarrow n\text{p}1_1$}& &
\multicolumn{2}{c}{$1\rightarrow n\text{p}1_2$}& &
\multicolumn{2}{c}{$3\rightarrow n\text{p}3_3$}& &
\multicolumn{2}{c}{$3\rightarrow n\text{p}3_4$}\\\cline{2-3}\cline{5-6}\cline{8-9}\cline{11-12}

\multicolumn{1}{l}{$n$}&
\multicolumn{1}{c}{$\tilde{\nu}_\text{obs}$}&
\multicolumn{1}{c}{$\Delta$}& &
\multicolumn{1}{c}{$\tilde{\nu}_\text{obs}$}&
\multicolumn{1}{c}{$\Delta$}& &
\multicolumn{1}{c}{$\tilde{\nu}_\text{obs}$}&
\multicolumn{1}{c}{$\Delta$}& &
\multicolumn{1}{c}{$\tilde{\nu}_\text{obs}$}&
\multicolumn{1}{c}{$\Delta$}\\ \hline

 97 &    34289.52624 &   -0.00004 &&   34289.56234 &    0.00033 &&   34284.65092 &   -0.00025 &&   34284.58058 &   -0.00019 \\
 98 &    34289.76334 &   -0.00025 &&   34289.79496 &    0.00012 &&   34284.88850 &    0.00002 &&   34284.79930 &   -0.00011 \\
 99 &    34289.99361 &   -0.00013 &&   34290.02153 &    0.00034 &&   34285.11883 &    0.00020 &&   34285.01232 &   -0.00022 \\
100 &    34290.21709 &    0.00008 &&   34290.24141 &    0.00032 &&   34285.34180 &   -0.00010 &&               &            \\
101 &    34290.43368 &   -0.00000 &&   34290.45530 &    0.00063 &&   34285.55861 &    0.00004 &&               &            \\
102 &    34290.64409 &    0.00008 &&   34290.66249 &    0.00041 &&   34285.76919 &    0.00029 &&               &            \\
103 &    34290.84840 &    0.00017 &&   34290.86374 &    0.00029 &&   34285.97326 &    0.00014 &&   34286.03132 &   -0.00045 \\
104 &    34291.04669 &    0.00010 &&   34291.05938 &    0.00025 &&   34286.17132 &   -0.00016 &&   34286.22484 &    0.00063 \\
105 &    34291.23961 &    0.00031 &&   34291.24928 &    0.00093 &&   34286.36407 &   -0.00012 &&   34286.41241 &    0.00032 \\
106 &    34291.42671 &    0.00013 &&   34291.43364 &    0.00136 &&   34286.55146 &   -0.00001 &&   34286.59561 &    0.00027 \\
107 &                &            &&               &            &&   34286.73320 &   -0.00033 &&   34286.77426 &    0.00029 \\
108 &                &            &&               &            &&   34286.90993 &   -0.00061 &&   34286.94813 &    0.00010 \\
109 &    34291.95807 &   -0.00001 &&   34291.94902 &    0.00023 &&   34287.08215 &   -0.00055 &&   34287.11789 &    0.00027 \\
110 &    34292.12479 &   -0.00055 &&   34292.10716 &   -0.00053 &&   34287.24973 &   -0.00046 &&   34287.28364 &    0.00080 \\
111 &    34292.28799 &   -0.00031 &&   34292.25790 &   -0.00048 &&   34287.41291 &   -0.00025 &&   34287.44433 &    0.00054 \\
112 &    34292.44664 &   -0.00028 &&   34292.39993 &   -0.00038 &&               &            &&               &            \\\hline

\\[-1.8ex] & \multicolumn{2}{c}{$5\rightarrow n\text{p}5_5$}& &
\multicolumn{2}{c}{$5\rightarrow n\text{p}5_6$}& &
\multicolumn{2}{c}{$7\rightarrow n\text{p}7_7$}& &
\multicolumn{2}{c}{$7\rightarrow n\text{p}7_8$}\\\cline{2-3}\cline{5-6}\cline{8-9}\cline{11-12}

\multicolumn{1}{l}{$n$}&
\multicolumn{1}{c}{$\tilde{\nu}_\text{obs}$}&
\multicolumn{1}{c}{$\Delta$}& &
\multicolumn{1}{c}{$\tilde{\nu}_\text{obs}$}&
\multicolumn{1}{c}{$\Delta$}& &
\multicolumn{1}{c}{$\tilde{\nu}_\text{obs}$}&
\multicolumn{1}{c}{$\Delta$}& &
\multicolumn{1}{c}{$\tilde{\nu}_\text{obs}$}&
\multicolumn{1}{c}{$\Delta$}\\\hline

 95 &    34275.39957 &    0.00004 &&   34275.42733 &    0.00011 &&   34262.80252 &    0.00009 &&   34262.80698 &   -0.00010 \\
 96 &    34275.65156 &   -0.00052 &&   34275.67834 &    0.00045 &&               &            &&               &            \\
 97 &    34275.89692 &    0.00007 &&   34275.92123 &    0.00036 &&   34263.29979 &    0.00005 &&               &            \\
 98 &    34276.13404 &   -0.00012 &&   34276.15671 &    0.00022 &&               &            &&               &            \\
 99 &    34276.36393 &   -0.00037 &&   34276.38536 &    0.00033 &&               &            &&               &            \\
100 &    34276.58738 &   -0.00020 &&   34276.60701 &    0.00025 &&               &            &&               &            \\
101 &    34276.80383 &   -0.00042 &&   34276.82230 &    0.00034 &&               &            &&               &            \\
102 &    34277.01420 &   -0.00037 &&   34277.03116 &    0.00027 &&   34264.41782 &    0.00039 &&   34264.40766 &    0.00047 \\
103 &    34277.21790 &   -0.00090 &&   34277.23394 &    0.00014 &&   34264.62158 &   -0.00004 &&   34264.60938 &    0.00016 \\
104 &                &            &&   34277.43131 &    0.00009 &&   34264.81930 &   -0.00030 &&   34264.80533 &   -0.00002 \\
105 &                &            &&   34277.62236 &    0.00024 &&               &            &&               &            \\
106 &                &            &&   34277.80848 &    0.00029 &&   34265.20041 &    0.00032 &&   34265.18098 &    0.00019 \\
107 &    34277.97814 &   -0.00106 &&   34277.98926 &    0.00021 &&   34265.38188 &   -0.00024 &&   34265.36024 &   -0.00025 \\
108 &    34278.15511 &   -0.00110 &&   34278.16509 &    0.00020 &&   34265.55876 &   -0.00037 &&   34265.53486 &   -0.00025 \\
109 &                &            &&   34278.33671 &    0.00080 &&   34265.73158 &    0.00029 &&   34265.70451 &   -0.00034 \\
110 &                &            &&   34278.50342 &    0.00114 &&               &            &&   34265.86985 &   -0.00003 \\\hline

\\[-1.8ex] & \multicolumn{2}{c}{$9\rightarrow n\text{p}9_{9}$}& &
\multicolumn{2}{c}{$9\rightarrow n\text{p}9_{10}$}& &
\multicolumn{2}{c}{$11\rightarrow n\text{p}11_{11}$}& &
\multicolumn{2}{c}{$11\rightarrow n\text{p}11_{12}$}\\\cline{2-3}\cline{5-6}\cline{8-9}\cline{11-12}

\multicolumn{1}{l}{$n$}&
\multicolumn{1}{c}{$\tilde{\nu}_\text{obs}$}&
\multicolumn{1}{c}{$\Delta$}& &
\multicolumn{1}{c}{$\tilde{\nu}_\text{obs}$}&
\multicolumn{1}{c}{$\Delta$}& &
\multicolumn{1}{c}{$\tilde{\nu}_\text{obs}$}&
\multicolumn{1}{c}{$\Delta$}& &
\multicolumn{1}{c}{$\tilde{\nu}_\text{obs}$}&
\multicolumn{1}{c}{$\Delta$}\\ \hline

 95 &                &            &&               &            &&   34226.28484 &   -0.00040 &&   34226.32783 &    0.00028 \\
 96 &                &            &&               &            &&   34226.53770 &   -0.00007 &&   34226.57889 &    0.00052 \\
 97 &                &            &&               &            &&   34226.78251 &   -0.00003 &&   34226.82174 &    0.00023 \\
 98 &                &            &&               &            &&   34227.01952 &   -0.00032 &&   34227.05785 &    0.00057 \\
 99 &                &            &&               &            &&   34227.25019 &    0.00021 &&   34227.28585 &   -0.00012 \\
100 &                &            &&               &            &&               &            &&               &            \\
101 &    34247.81132 &   -0.00084 &&   34247.79465 &   -0.00056 &&   34227.69002 &    0.00010 &&   34227.72374 &    0.00052 \\
102 &    34248.02325 &   -0.00011 &&   34248.00394 &   -0.00050 &&   34227.90013 &   -0.00011 &&   34227.93272 &    0.00042 \\
103 &    34248.22707 &   -0.00044 &&   34248.20704 &    0.00086 &&   34228.10335 &   -0.00111 &&   34228.13595 &    0.00061 \\
104 &                &            &&   34248.40210 &   -0.00001 &&   34228.30168 &   -0.00113 &&   34228.33351 &    0.00094 \\
105 &                &            &&   34248.58708 &   -0.00317 &&   34228.49511 &   -0.00041 &&   34228.52468 &    0.00047 \\
106 &    34248.80635 &    0.00053 &&   34248.78080 &   -0.00030 &&   34228.68271 &   -0.00009 &&               &            \\
107 &    34248.98751 &   -0.00035 &&   34248.96020 &   -0.00014 &&   34228.86415 &   -0.00069 &&   34228.89115 &   -0.00041 \\
108 &                &            &&   34249.13606 &    0.00118 &&   34229.04169 &   -0.00016 &&               &            \\
109 &                &            &&   34249.30431 &   -0.00073 &&               &            &&               &            \\
110 &    34249.50550 &    0.00099 &&   34249.47092 &    0.00030 &&               &            &&   34229.40579 &    0.00023 \\
\hline\hline
\multicolumn{12}{r}{continued on next page}
\end{tabular}
\end{table*}
\endgroup

\addtocounter{table}{-1}

\begingroup
\squeezetable
\begin{table*}[t!]
\caption{
--\textit{Continued}}
\begin{tabular}{
s[table-unit-alignment = left,table-format=3]
S[table-format=6.5]
S[table-format=-2.5]
c
S[table-format=6.5]
S[table-format=-2.5]
c
S[table-format=6.5]
S[table-format=-2.5]
c
S[table-format=6.5]
S[table-format=-2.5]}

\hline\hline
& \multicolumn{2}{c}{$13\rightarrow n\text{p}13_{13}$}& &
\multicolumn{2}{c}{$13\rightarrow n\text{p}13_{14}$}& &
\multicolumn{2}{c}{$15\rightarrow n\text{p}15_{15}$}& &
\multicolumn{2}{c}{$15\rightarrow n\text{p}15_{16}$}\\\cline{2-3}\cline{5-6}\cline{8-9}\cline{11-12}

\multicolumn{1}{l}{$n$}&
\multicolumn{1}{c}{$\tilde{\nu}_\text{obs}$}&
\multicolumn{1}{c}{$\Delta$}& &
\multicolumn{1}{c}{$\tilde{\nu}_\text{obs}$}&
\multicolumn{1}{c}{$\Delta$}& &
\multicolumn{1}{c}{$\tilde{\nu}_\text{obs}$}&
\multicolumn{1}{c}{$\Delta$}& &
\multicolumn{1}{c}{$\tilde{\nu}_\text{obs}$}&
\multicolumn{1}{c}{$\Delta$}\\ \hline

 95 &                &            &&               &            &&   34175.18430 &   -0.00085 &&   34175.23264 &   -0.00070 \\
 96 &    34202.76638 &   -0.00054 &&   34202.81975 &   -0.00030 &&   34175.43746 &   -0.00022 &&               &            \\
 97 &    34203.01109 &   -0.00058 &&   34203.06227 &   -0.00039 &&   34175.68158 &   -0.00085 &&   34175.72809 &    0.00102 \\
 98 &    34203.24813 &   -0.00083 &&   34203.29739 &   -0.00055 &&   34175.91962 &   -0.00011 &&   34175.96313 &    0.00041 \\
 99 &    34203.47855 &   -0.00054 &&   34203.52625 &    0.00008 &&   34176.14954 &   -0.00033 &&   34176.19244 &    0.00114 \\
100 &    34203.70153 &   -0.00082 &&   34203.74780 &    0.00017 &&               &            &&   34176.41276 &   -0.00032 \\
101 &    34203.91840 &   -0.00061 &&   34203.96297 &    0.00038 &&   34176.59000 &    0.00021 &&   34176.62850 &    0.00017 \\
102 &    34204.12953 &    0.00021 &&   34204.17203 &    0.00074 &&   34176.79939 &   -0.00071 &&               &            \\
103 &    34204.33355 &    0.00002 &&               &            &&               &            &&               &            \\
104 &    34204.53344 &    0.00156 &&   34204.57122 &    0.00034 &&   34177.20282 &    0.00016 &&               &            \\
105 &    34204.72462 &    0.00004 &&   34204.76464 &    0.00243 &&   34177.39419 &   -0.00118 &&   34177.42973 &    0.00082 \\
106 &    34204.91172 &   -0.00013 &&   34204.94854 &    0.00037 &&   34177.58125 &   -0.00139 &&   34177.61478 &   -0.00030 \\
107 &    34205.09357 &   -0.00033 &&   34205.12912 &    0.00015 &&   34177.76308 &   -0.00160 &&   34177.79600 &   -0.00006 \\
108 &    34205.27096 &    0.00006 &&               &            &&               &            &&               &            \\
109 &    34205.44272 &   -0.00034 &&   34205.47635 &    0.00053 &&               &            &&               &            \\
110 &    34205.61031 &   -0.00022 &&               &            &&   34178.28090 &   -0.00042 &&               &            \\\hline

\\[-1.8ex] & \multicolumn{2}{c}{$17\rightarrow n\text{p}17_{17}$}& &
\multicolumn{2}{c}{$17\rightarrow n\text{p}17_{18}$}& &
\multicolumn{2}{c}{$19\rightarrow n\text{p}19_{19}$}& &
\multicolumn{2}{c}{$19\rightarrow n\text{p}19_{20}$}\\\cline{2-3}\cline{5-6}\cline{8-9}\cline{11-12}

\multicolumn{1}{l}{$n$}&
\multicolumn{1}{c}{$\tilde{\nu}_\text{obs}$}&
\multicolumn{1}{c}{$\Delta$}& &
\multicolumn{1}{c}{$\tilde{\nu}_\text{obs}$}&
\multicolumn{1}{c}{$\Delta$}& &
\multicolumn{1}{c}{$\tilde{\nu}_\text{obs}$}&
\multicolumn{1}{c}{$\Delta$}& &
\multicolumn{1}{c}{$\tilde{\nu}_\text{obs}$}&
\multicolumn{1}{c}{$\Delta$}\\\hline

 95 &    34144.41006 &    0.00041 &&   34144.44231 &    0.00018 &&               &            &&   34110.27532 &    0.00158 \\
 96 &    34144.66117 &   -0.00101 &&   34144.69419 &    0.00075 &&   34110.55135 &   -0.00008 &&   34110.52419 &   -0.00112 \\
 97 &    34144.90604 &   -0.00089 &&   34144.93709 &    0.00007 &&   34110.79683 &    0.00068 &&               &            \\
 98 &    34145.14392 &   -0.00031 &&   34145.17317 &   -0.00003 &&   34111.03332 &   -0.00011 &&   34111.00606 &    0.00039 \\
 99 &    34145.37291 &   -0.00145 &&   34145.40204 &   -0.00024 &&   34111.26453 &    0.00098 &&   34111.23489 &   -0.00017 \\
100 &    34145.59743 &   -0.00019 &&   34145.62402 &   -0.00051 &&   34111.48817 &    0.00137 &&   34111.45759 &   -0.00003 \\
101 &    34145.81464 &    0.00036 &&               &            &&   34111.70224 &   -0.00121 &&               &            \\
102 &    34146.02488 &    0.00029 &&   34146.04952 &   -0.00011 &&               &            &&               &            \\
103 &    34146.22890 &    0.00010 &&   34146.25415 &    0.00118 &&               &            &&   34112.08648 &   -0.00052 \\
104 &                &            &&               &            &&   34112.31775 &    0.00145 &&               &            \\
105 &                &            &&               &            &&   34112.50860 &   -0.00040 &&               &            \\
106 &    34146.80723 &    0.00012 &&               &            &&   34112.69634 &    0.00008 &&   34112.66306 &   -0.00088 \\
107 &    34146.98807 &   -0.00108 &&               &            &&   34112.87955 &    0.00125 &&   34112.84505 &   -0.00055 \\
108 &    34147.16622 &    0.00006 &&   34147.18699 &    0.00048 &&   34113.05443 &   -0.00087 &&   34113.02226 &   -0.00001 \\
109 &    34147.33784 &   -0.00047 &&   34147.35928 &    0.00128 &&   34113.22723 &   -0.00022 &&   34113.19394 &   -0.00018 \\
110 &    34147.50469 &   -0.00109 &&   34147.52605 &    0.00121 &&   34113.39480 &   -0.00012 &&               &            \\
\hline\hline
\end{tabular}
\end{table*}
\endgroup

To obtain a full set of rotational intervals in the $a\,^3\Sigma_u^+$ state of He$_2$ and the $X^+\,^2\Sigma_u^+$ state of He${_2}^+$, we proceeded in two steps. First, initial-state combination differences $\tilde{\nu}_{N''-2,N''}$, i.e., the energy intervals between the $N''-2$ and $N''$ rotational levels, were determined as illustrated in Fig.~\ref{fig:Olines} with the example of the $N''=15$ and 17 levels, where the upper and lower panels show the $n\text{p}15_{16}$ series around $n=57$ recorded from the $N''=15$ and $N''=17$ levels, respectively. Each member of this series observed from both rotational levels represents an independent measurement of $\tilde{\nu}_{15,17}$. The differences between the centers of gravity of the triplets (marked by vertical dashed lines) give a value of \num{481.1061\pm0.0007}\,cm$^{-1}$ for the interval between the $a\,^3\Sigma_u^+,\,\nu''=0,N''=15$ and 17 levels. All transitions used to determine the rotational structure of the $a\,^3\Sigma_u^+$ state are listed in Table~S1 of the Supplementary Material. These transitions form a redundant network, from which the relative positions of all states connected by the transitions can be determined in a global fit, as described in Ref.~\cite{schafer2010}. The rotational term values of the $a\,^3\Sigma_u^+\,(\nu''=0)$ metastable state of He$_2$ presented in the third column of Table~\ref{tab:compare_focsa} could be determined in this way up to $N''=19$. These term values agree, within the respective uncertainties, with the rotational term values we calculated with the rotational constants reported by Focsa \emph{et al.} \cite{focsa1998} and presented in the second column of Table~\ref{tab:compare_focsa}. 

In the second step, the positions of the rotational levels of the He${_2}^+$ ion were determined by extrapolating $n\text{p}[N^+=N'']_{N^+,N^+=1}\leftarrow N''$ series. Figure~\ref{fig:Qseries} presents spectra of the $n\text{p}[N^+=N'']_{N=N'',N''+1}\leftarrow N''$ Rydberg series recorded for this purpose, and Table~\ref{tab:Qlines} lists the corresponding transition wavenumbers. The choice of the range of principal quantum number between 95 and 110 used for the extrapolation represented the best compromise between maximal accuracy of the Rydberg-series extrapolation, which is reached at the highest possible values of $n$, and minimal systematic uncertainties in the wavenumber measurements. In our experiments  (see discussion below), the dominant source of systematic uncertainty comes from Stark-shifts induced by residual electric fields, which could be compensated to 1\,mV/cm. This uncertainty amounts to 10\,MHz at $n=110$ and scales with $n^7$, which made extrapolations based on transitions including Rydberg states with $n>110$ insufficiently accurate. 

The spectra depicted in Fig.~\ref{fig:Qseries} all reveal at least one of the two expected series. In some spectra, such as the spectrum of the $n\text{p}1_{1,2}\leftarrow 1$ transitions, perturbations lead to regions where the two series are clearly distinguishable (e.g., in the region $n=100-105$ and above $n=109$ in the $n\text{p}1_{1,2}$ series). In other spectra, the two series are difficult to distinguish because of the partially overlapping fine structure and reliable ``fine-structure-free'' positions can only be derived from the analysis of lineshapes. 

In an initial fit, the $n\text{p}19_{20}\leftarrow 19$ series was excluded, because of its interaction with the $n\text{p}21_{20}$ series which could not be measured at sufficiently high $n$ values for the $N^+=21$ threshold to be accurately determined. The rotational energy levels with $N^+\le 19$ obtained in this way were used to derive a set of effective rotational constants that was subsequently used to predict the term value of the $N^+=21$ level with respect to the $N^+=1$ level (3158.115\,cm$^{-1}$). The ionization thresholds of all series, including the $n\text{p}19_{20}$ series, were then determined in a second fit for which the $N^+=21$ threshold was fixed at the predicted value. The relative positions of the ionic rotational states obtained from the transition wavenumbers listed in Table~\ref{tab:Qlines} and from the known positions of the rotational levels of the metastable state (see third column of Table~\ref{tab:compare_focsa}) are listed in the last column of Table~\ref{tab:compare_focsa}, where they are compared with the results of the \emph{ab initio} calculations of Tung \emph{et al.} \cite{tung2012}. The adiabatic ionization energy of He$_2$ ($a\,^3\Sigma_u^+$), defined as the interval between the $a\,^3\Sigma_u^+\,(\nu''=0,N''=1)$ state of He$_2$ and the $X^+\,^2\Sigma_u^+\,(\nu^+=0,N^+=1)$ state of He${_2}^+$, is found to be $\num{34301.20565\pm0.00012}\pm0.0014_\text{sys}$\,cm$^{-1}$, as already reported in Ref.~\cite{jansen2015}.

\begin{figure}[bt]
\centering
\includegraphics[width=.9\columnwidth]{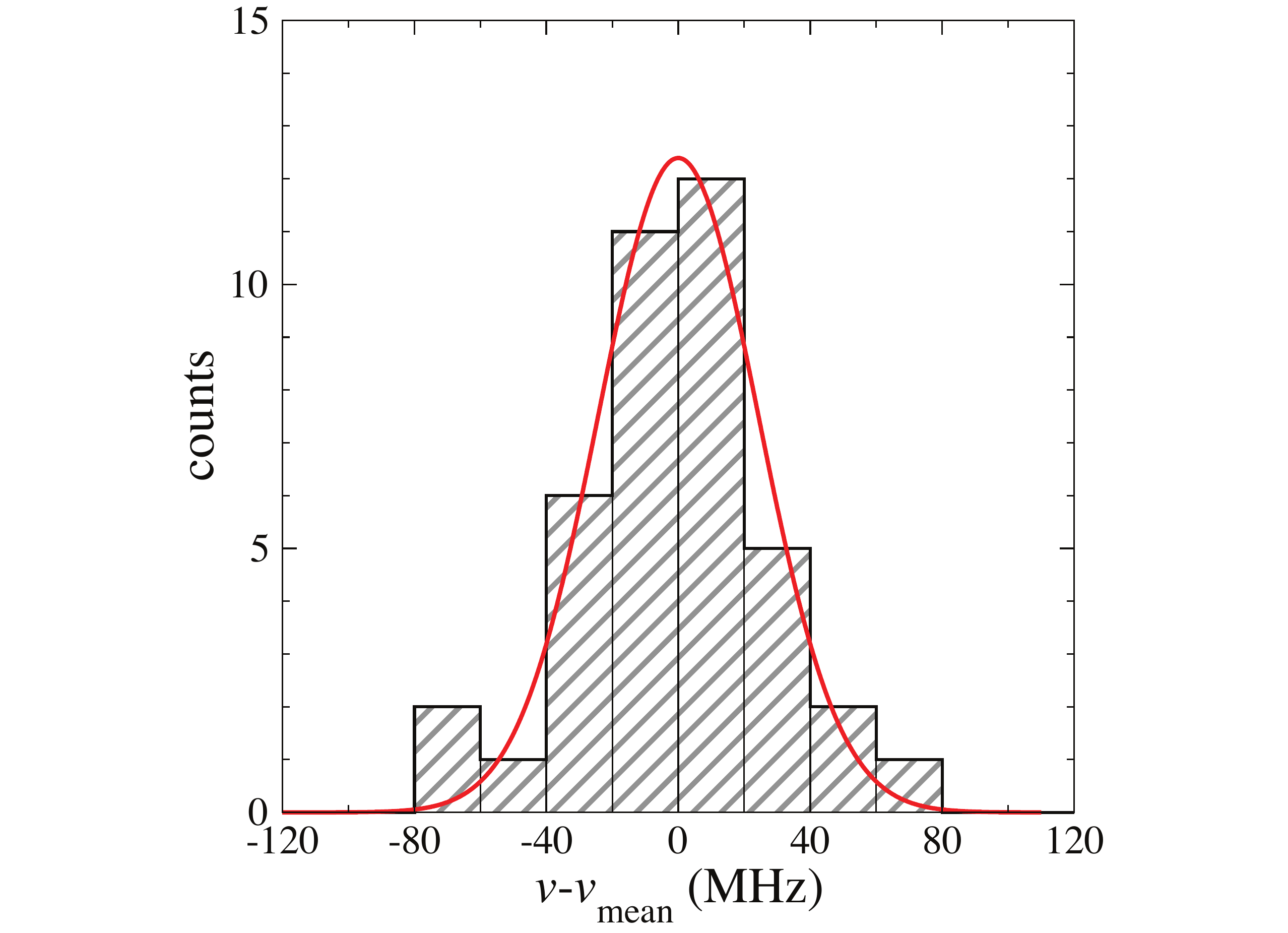}
\caption{Histogram to quantify the relative accuracy of the wavemeter using the distribution of transition wavenumbers of the $124\text{p}1_2\leftarrow 1$ transition determined in independent calibration spectra recorded before each measurement. The solid line represents a Gaussian fit to the data with a standard deviation of 25\,MHz. }
\label{fig:wavemeter_hist}
\end{figure}

Table~\ref{tab:error_budget} summarizes the main sources of systematic and statistical uncertainty in the determination of ionization thresholds and He${_2}^+$ rotational intervals. The main source of statistical uncertainty comes from the lineshape analysis and the wavenumber calibration (25\,MHz), resulting from the accuracy of the wavemeter. The systematic uncertainty of the ionization thresholds is larger (40\,MHz) than that of the rotational intervals because the absolute calibration error cancels out when building combination differences. Although the absolute accuracy of the wave-meter measurements is specified as 20\,MHz, the intrinsic Allan deviation over periods shorter than 1 min lies typically around 300\,kHz \cite{sanguinetti2009,deiglmayr2016}. Long-term fluctuations of the wavemeter were monitored by recording a small part of the spectrum prior to each measurement and performing a statistical analysis of the fitted line positions. The resulting histogram is shown in Fig.~\ref{fig:wavemeter_hist} and resembles a Gaussian distribution with a standard deviation of 25(5)\,MHz in the UV. This uncertainty of 25\,MHz approximately corresponds to the specified relative accuracy of the wavemeter of 10\,MHz in the fundamental and contributes a statistical uncertainty of 25\,MHz to the rotational intervals. The uncertainty of the Rydberg-series extrapolation by MQDT is dominated by the uncertainty in the quantum-defect parameters reported in Ref.~\cite{sprecher2014} and is only 3\,MHz for the ionization energy and 6\,MHz for the rotational intervals. This uncertainty does not cancel out when determining the relative positions of the ionic levels because the series of different $N$ values were extrapolated separately. 

The remaining systematic uncertainties are dominated by the contribution of the Stark shift resulting from the residual electric field. The other sources of systematic uncertainties only affect the value of the ionization energy because they cancel out when determining the relative positions of the ionic rotational levels. 

As explained in Sec.~\ref{sec:experiment}, the stray electric field was determined and subsequently compensated by recording Rydberg spectra in the presence of different dc electric-field strengths and fitting the observed Stark shifts to a quadratic polynomial. Stray magnetic fields were suppressed by surrounding the photoexcitation region by two concentric mu-metal shields. The contribution of Stark and Zeeman shifts induced by residual electric and magnetic fields are 10 and 1\,MHz, respectively. No shift in the line positions was observed when the laser power and the nozzle-stagnation pressure were varied, and the contributions of ac-Stark and pressure shifts to the systematic uncertainties is estimated to be at most 2 and 1\,MHz, respectively. Uncertainties from Doppler shifts were determined to be around 5\,MHz.

\begingroup
\begin{table}
\caption{\label{tab:error_budget}
Estimated statistical ($1\sigma$) and systematic contributions to the uncertainty in the determinations of the ionization threshold of the $a\,^3\Sigma_u^+$ state of He$_2$ (second column) and rotational intervals of the He${_2}^+\,X^+\,^2\Sigma_u^+,\,\nu^+=0$ ground state (third column). All values are given in MHz.}
\begin{tabularx}{\columnwidth}{@{\extracolsep{\fill}}l l c@{\hspace{40pt}}c @{}}
\toprule
\multicolumn{2}{l}{Source}&
$E_\text{I}$&
$\tilde{\nu}_{N^++2,N^+}$\\
\hline
\multicolumn{4}{l}{\emph{Statistical}}\\
 & Wavenumber calibration	 &  0 & 25\\
 & Determination of line centers & 2  & 3\\
 & MQDT fit			 &   3 &  5-8\\
\cline{2-4}
 & Total statistical\bigstrut    &  4 & 26 \\
\hline
\multicolumn{4}{l}{\emph{Systematic}}\\
 & Wavenumber calibration	 &  40 & 0\\
 & AC-Stark shift 		 &2 	&0 \\
 & DC-Stark shift 		 &10	&5 \\
 & Zeeman shift   		 &1 	&0 \\
 & Doppler shift 		 &5 	&0 \\
 & Pressure shift 		 &1 	&0 \\
\cline{2-4}
 & Total systematic\bigstrut     &   42 & 5 \\
\toprule

\end{tabularx}
\end{table}

\endgroup

\section{Discussion and conclusions\label{sec:conclusion}}
The most precise \emph{ab initio} calculations of the rovibrational energy levels of He${_2}^+$ reported so far have been carried out by Tung \emph{et al.} \cite{tung2012}. The calculations relied on the use of 1200 explicitly correlated Gaussian functions with shifted centers in the range of internuclear distances between $R=0.9$ and $100\,a_0$ to obtain the Born-Oppenheimer potential energy function. To derive accurate rovibrational energies, Tung \emph{et al.} \cite{tung2012} also calculated the adiabatic corrections and included nonadiabatic corrections by using $R$-independent effective reduced masses in the vibrational and rotational kinetic operators in the Hamiltonian. Relativistic and radiative corrections were neglected, but using an argument based on scaling these corrections from the values they take in H$_2$ \cite{komasa2011} to the relativistic correction for the first vibrational interval of $^3$He$^4$He$^+$ \cite{stanke2009} and assuming that these corrections are mass independent, they estimated their predictions of the lowest vibrational interval in He${_2}^+$ to have an accuracy of about 0.004\,cm$^{-1}$, or 120\,MHz. 

\begin{figure}[th]
\centering
\includegraphics[width=1\columnwidth]{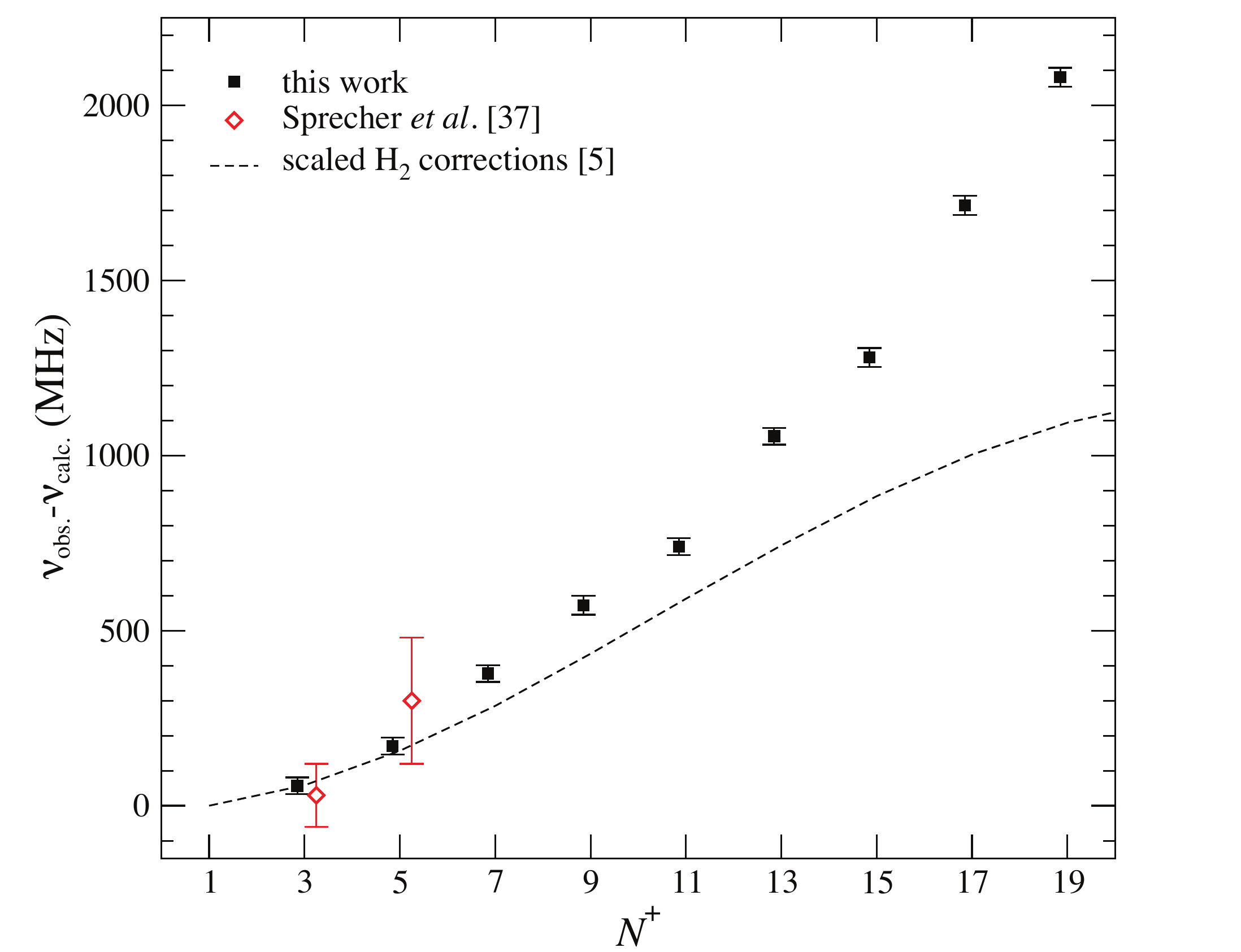}
\caption{Differences between experimental (this work) and theoretical rotational term values (from Ref.~\cite{tung2012}) in the $X^+\,^2\Sigma_u^+\,(\nu^+=0)$ state of He${_2}^+$ (full squares). The experimental results of Sprecher \emph{et al.} \cite{sprecher2014} are plotted as diamonds. The dashed line represents the corrections to the energy levels of molecular hydrogen that were scaled to the relativistic correction of the fundamental vibration in $^3$He$^4$He$^+$ (see text).}
\label{fig:obs-calc}
\end{figure}

Comparison of our experimental results with the predictions of Tung \emph{et al.} \cite{tung2012}, we find that the experimental term values always lie above the calculated ones, and that the discrepancies rapidly increase with increasing rotational quantum number (see Fig.~\ref{fig:obs-calc}), reaching a value of almost 0.07\,cm$^{-1}$ (2.1\,GHz) at $N^+=19$, which is more than ten times larger than the estimated uncertainty of the calculations. In reconstructing the argument of Tung \emph{et al.} \cite{tung2012}, we found that the relativistic \mbox{(-0.0026}\,cm$^{-1}$) and radiative \mbox{(0.0010}\,cm$^{-1}$) corrections of the first rotational interval of H$_2$ \cite{komasa2011}, rather than the first vibrational interval, were used to scale the relativistic correction of the first vibrational interval in $^3$He$^4$He$^+$ (-0.006 cm$^{-1}$) \cite{stanke2009}. When we scaled the relativistic \mbox{(-0.0234}\,cm$^{-1}$) and radiative (0.0213\,cm$^{-1}$) corrections of the $(\nu,J):(0,0)\rightarrow(1,0)$ interval in H$_2$ \cite{komasa2011}, we found a correction of 0.0005\,cm$^{-1}$, or 15\,MHz, for this interval in $^3$He$^4$He$^+$. Following the argument of Tung \emph{et al.}, we scaled the relativistic and radiative corrections of the rotational states of H$_2$ \cite{komasa2011} by the factor 3.9. The results are displayed as the dashed line in Fig.~\ref{fig:obs-calc}. The scaled corrections appear to account for the discrepancies between experimental and calculated rotational term values reasonably well up to $N^+=11$, but rapidly become inadequate beyond $N^+=11$. Further work is required to understand the origin of this discrepancy. It is conceivable that the approximate treatment of the nonadiabatic correction through $R$-independent reduced masses in Ref.~\cite{tung2012} becomes increasingly inaccurate as $N^+$ increases. 

The results presented in this article suggest that even the best \emph{ab initio} calculations available in the literature on the three-electron molecule He${_2}^+$ are still far from reaching the same level of agreement with experimental data as in the two-electron molecule H$_2$ \cite{liu2009,piszczatowski2009,komasa2011,salumbides2011,dickenson2013}.

\section*{Supplementary Material}
See Supplementary Material at [address to be inserted by Editor] for the wavenumbers of all transitions used to derive the rotational level energies of the $a\,^3\Sigma_u^+\,(\nu''=0)$ state of He${_2}$.

\begin{acknowledgments}
We thank Christian Jungen (Laboratoire Aimé Cotton du CNRS, Orsay) for allowing us to use his MQDT program and Hansjürg Schmutz and Josef Agner for their expert technical assistance. This work is supported financially by the Swiss National Science Foundation under Project Nos. 200020-159848 and 200020-149216 and the NCCR QSIT. P.J. gratefully acknowledges ETH Zurich for support through an ETH fellowship.
\end{acknowledgments}

%

\end{document}